\RequirePackage{fix-cm}
\documentclass[aps,prd,showpacs,showkeys,superscriptaddress,twocolumn]{revtex4-1}
\usepackage[colorlinks,linkcolor=blue,anchorcolor=blue,citecolor=blue,urlcolor=blue,breaklinks=true]{hyperref}
\usepackage[T1]{fontenc}
\usepackage{graphicx}% Include figure files
\usepackage{amsmath}
\usepackage{amssymb}
\usepackage{slashed}
\usepackage{latexsym}
\usepackage{epsfig}
\usepackage{amsbsy}
\usepackage{array}
\usepackage{amssymb}
\usepackage{setspace}
\usepackage{bm}
\usepackage{lipsum}
\usepackage{mathrsfs}
\usepackage{float}
\usepackage{changes}
\usepackage{color}
\usepackage{booktabs}
\usepackage{mathptmx}
\usepackage{graphicx}
\usepackage{epstopdf}
\newcommand{\Lagr}{\mathcal{L}}
\newcommand{\fie}{\mathcal{F}}
\begin{document}
	\author{Bing-Jun Zuo}
	\email{812380695@qq.com}
	\affiliation{Department of Physics, Nanjing University, Nanjing 210093, China}
	\author{Yong-Feng Huang}
	\email{hyf@nju.edu.cn}
	\affiliation{School of Astronomy and Space Science, Nanjing University, Nanjing 210023, China}
	\affiliation{Key Laboratory of Modern Astronomy and Astrophysics (Nanjing University), Ministry of Education, Nanjing 210023, China}
	\author{Hong-Tao Feng}
	\affiliation{School of Physics, Southeast University, Nanjing 211189,  China}
	\date{\today}
	
	\title{Hybrid stars and QCD phase transition with a NJL-like model}
	
	\begin{abstract}
		In this paper, we introduce a  self-consistent  mean field approximation to study the QCD phase transition and the structure of hybrid stars within the framework of  NJL model. In our practice, a phenomenological parameter $\alpha$ is introduced, which reflects the weights of ``direct'' channel and ``exchange'' channel under a finite chemical potential.  The mass-radius relation is obtained by solving the Tolman-Oppenheimer-Volkoff  equation  using a crossover equation of state (EOS). We calculate the density distribution in a two solar-mass hybrid star to show the effects of different parameters.  We also calculate the tidal Love number $k_2$ and the  deformability $(\Lambda)$. It is found that the stiffness of the EOS   increases with  $\alpha $, which allows us to obtain a hybrid star  with a maximum mass of 2.40 solar-mass through our model. The observation of over 2.06 solar-mass neutron stars may indicates that the chiral transition may be a crossover on the whole $T-\mu$ plane.
		\bigskip
	\end{abstract}
	
	\flushbottom
	\maketitle
	\thispagestyle{empty}
	
	\section{Introduction}
	Neutron stars are the  densest celestial bodies in our universe except for  black holes. Their extreme environment provides us a natural laboratory for cold and condensed matter. Especially, in recent years,  the detection of gravitational waves has given us  different valuable information about compact stars. During the inspiral and merger of binary stars, the tidal deformability  can be measured  by the LIGO and VIRGO network directly \cite{abbott2017gw170817}. The Love number $k_2$ is used to measure the distortion of a neutron star. It can be calculated through  the exterior solutions that are related to the tidal deformability $(\Lambda)$ \cite{damour1992general,flanagan2008constraining,hinderer2008tidal}. On the other hand, electromagnetic observations on the cooling rate\cite{Dunne:2004nc}, gamma-ray bursts \cite{Berger:2013jza}, thermal X-rays\cite{Miller:2019cac} and radio bursts \cite{Geng:2021apl} from neutron stars give us diverse data to study  neutron stars. These astronomical observations can  help us to constrain theoretical models and improve our understanding of neutron stars. \\
	\indent   The inner structure of neutron stars is still not well known. The possibility that the neutron star is composed of a quark matter inner core, nucleon matter outer core and crust, i.e., being a hybrid star, is widely discussed.  The key to study  hybrid stars is the equation of state (EOS) of the strongly interacting matter. For quark matter,  lattice  quantum chromodynamics (LQCD) is widely accepted as the ab-initio method dealing with  QCD nonperturbatively. Unfortunately, present lattice QCD calculations at finite chemical potential are plagued with the so called ``sign problem''  \cite{bethke2007experimental}. Thus, to study QCD systems at finite chemical potential, it is necessary to employ some QCD effective models, such as the Nambu-Jona-Lasinio (NJL) model \cite{vogl1991nambu,klevansky1992nambu,hatsuda1994qcd,fan2017mapping,li2019new}.  The NJL model displays  important features of QCD, i.e., the dynamical chiral symmetry breaking and restoring. The chiral restoring phase transition is usually considered to happen simultaneously (or almost simultaneously) with the deconfinement phase transition \cite{Philipsen:2012nu,Fodor:2009ax,Gupta:2007ax,Borsanyi:2013bia}. Thus, a problem naturally arises: which type of phase transition should it be, a first order phase transition or a crossover? \\
	\indent Another related issue is that in the case of zero chemical potential and finite temperature, the result of LQCD and other methods shows there is a crossover\cite{Philipsen:2012nu,Fodor:2009ax,Gupta:2007ax,Borsanyi:2013bia}. Thus, at zero temperature and  finite chemical potential the type of phase transition determines whether there exists a critical end point (CEP) in the QCD phase diagram. Experimentally, the second stage beam energy scanning plan of Relativistic Heavy Ion Collider (RHIC)  is still to try to find the possible CEP \cite{aggarwal2010higher,adamczyk2014energy,adamczyk2018collision}. Theoretically, different models  give different perspectives and answers.  In this paper, we will study this issue by using a self-consistent mean field approximation in the framework of NJL model \cite{wang2019novel}.  People generally predict that the hadronic matter will completely transit to quark matter when the density increases to 4--8 times the saturation density of nucleon matter in the hadron physical pictures. \cite{baym1976can,baym2018hadrons,li2018constraints,doi:10.1063/1.5117820}, which means that the corresponding quark chemical potential is at least larger than 430 MeV in NJL model. However, the strong interaction phase transition predicted by the standard NJL model usually occurs near 330 MeV \cite{masayuki1989chiral,andersen2002equation,fortin2016neutron,li2018studies}, which indicates that there is an contradiction between the prediction of the standard NJL model and the corresponding prediction of the hadron physical pictures.\\
	\indent In order to overcome  this contradiction, the authors of Ref. \cite{wang2019novel}
	proposed a  mean field approximation method under the framework of  NJL model. In this method,
	a phenomenological parameter $\alpha$ is introduced to reflect the weights of the ``exchange''
	channel and ``direct'' channel, because the vector interaction term generated by ``exchange''
	channel plays a very important role in the case of  finite chemical potential.  The relative
	results will influence the CEP and hybrid stars.
	$\alpha$ is a theoretical parameter.
		The finite density experiments could help to constrain $\alpha$. For example, the
		experiments of RHIC can provide useful clues on the issue of whether CEP exists or not,
		which itself can help constrain the value of $\alpha$. Anyway, note that such a constraint
		will still be too rough to determine $\alpha$ exactly, since there may be other factors
		influencing the CEP. Thus we need various experiments and observations, such as the
		astronomical observations of neutron stars, to further constrain the parameter. 
	Another point we focus on is the uncertain region between the nucleon matter and the quark matter. The uncertainty forces us to adopt some non-physical
	methods to connect the nucleon part and the quark part, such as the Gibbs construction\cite{Chen:2011my}, or other mathematical
	operations \cite{Masuda:2012ed,baym2018hadrons,PhysRevC.104.L012801}. However, these methods unavoidably lead to a  phase transition themselves, thus they are not self-consistent in the practice. The EOS will be studied under a crossover-type model with a self-consistent  connection method in this paper.  \\
	\indent This paper is organized as follows. In Section \ref{s2}, we introduce the three flavor NJL model and its Fierz transformation. The three-flavor QCD phase diagram is calculated and  the existence of  CEP in our model is discussed.  In Section \ref{s3}, the thermodynamical relations will  be solved to obtain the EOS for quark matter. We use a soft hadronic EOS  (APR)\cite{Akmal:1998cf} and a crossover connection method to get the EOS of hybrid stars. We will show how hybrid stars are changed in this framework.  Finally, we give a summary  in Section \ref{s4}.

	\section{MEAN FIELD APPROXIMATION AND PHASE TRANSITION}\label{s2}

	\indent  The standard 2+1 flavor NJL Lagrangian contains four-fermion and six-fermion interactions written as
	\begin{align}
		\label{E1}\Lagr=&\bar{\psi}(\emph{i}\slashed{\partial}-m)\psi+G\sum\limits_{a=0}^8[(\bar{\psi}\lambda^a\psi)^2+(\bar{\psi}\emph{i}\gamma_5\lambda^a\psi)^2]\\{}&-K[\textrm{det}\bar{\psi}(1+\gamma_5)\psi+\textrm{det}\bar{\psi}(1-\gamma_5)\psi]+\mu\psi^\dagger\psi\notag ,
	\end{align}
	where $m$ is  the current quark mass, $G$ and $K$ denote the coupling constants, which will be calibrated to reproduce the physical pion meson mass,
	kaon meson mass and  their decay constant. The corresponding mean field approximation is
	\begin{align}
		\Lagr_1\approx &\bar{\psi}[\emph{i}\slashed{\partial}-\gamma^0\mu-(m+\boldsymbol{A})]\psi,
	\end{align}
	with
	\[ \boldsymbol{A} =-4G\left( \begin{array}{ccc}\sigma_u&\quad&\quad\\ \quad  &\sigma_d &\quad  \\ \quad&\quad   &\sigma_s\\
	\end{array} \right)+2K\left( \begin{array}{ccc}\sigma_d\sigma_s&\quad&\quad\\ \quad  &\sigma_u\sigma_s &\quad  \\ \quad&\quad   &\sigma_u\sigma_d\\
	\end{array} \right) \]
	and
	\begin{align}
		\sigma_i=<\bar{\psi}_i\psi_i>=-\int\frac{d^4p}{(2\pi)^4}\textrm{Tr}[s(p)_i]\notag \qquad i=u, d, s ,
	\end{align}
	where $\sigma$ is the quark condensation, Tr denotes a trace over color and spinor  indices. \\
	\indent As shown in Eq. (\ref{E1}), the standard NJL model Lagrangian contains only the interactions of scalar and pseudoscalar channels. It is insufficient to handle the interaction of vector channels, such as in the case of finite chemical potential. To get a self-consistent result in the sense of mean field approximation, the contribution of Fierz transformation Lagrangian must be taken into account \cite{hatsuda1985soft}. The authors of Ref. \cite{klevansky1992nambu} demonstrated that a Fierz transformation of six-fermion interaction can be defined as an  operation that leaves the interaction invariant under all possible permutations of the quark spinors $\psi$ occurring in it, thus the six-fermion term dose not change after Fierz transformation. So in this paper we only need to consider the Fierz transformation of four-fermion term, which can be written as \cite{klevansky1992nambu}:
	\begin{align}\label{eq:L2}
		\fie\Biggl[\sum\limits_{a=0}^8[(\bar{\psi}\lambda^a\psi)^2+(\bar{\psi}\emph{i}\gamma_5\lambda^a\psi)^2]\Biggr]\notag\\=-\frac{G}{2}\sum\limits_{c=0}^8[(\bar{\psi}\gamma^\mu\lambda^c\psi)^2-(\bar{\psi}\gamma^\mu\gamma^5\lambda^c\psi)^2] .
	\end{align}
	\indent In this study, we only consider the scalar and vector channel contribution, this is because other terms in our modeling make no contribution at the level of mean field approximation. The mean field approximation of the Fierz transformation of the original NJL Lagrangian is
	\begin{align}
		\Lagr_2\approx &\bar{\psi}[\emph{i}\slashed{\partial}-\gamma^0(\mu+\boldsymbol{B})-(m+\boldsymbol{A})]\psi.
	\end{align}
	Here $\boldsymbol{B} =-\frac{2G}{3}(n_u+n_d+n_s)\boldsymbol{I},  \boldsymbol{I}$ is the identity matrix in flavor space and $n$ denotes the quark number density. $\Lagr_1$ contains only the Hartree term, while $\Lagr_2$ contains only the Fock term. We can rewrite Eq. (\ref{E1}) by taking the linear combination of $\Lagr_1$ and $\Lagr_2$ \cite{wang2019novel}:
	\begin{align}\label{E5}
		\Lagr_R=(1-\alpha)\Lagr_1+\alpha\Lagr_2.
	\end{align}
	\indent To illustrate the characteristics of this approach, let us recall the mean field approximation utilized in  previous studies within the framework of NJL model. In the original mean field approximation, the Fierz transformation is not adopted, which corresponds to the case of $\alpha=0$ in our method \cite{Nambu:1961tp}. But as pointed out in Refs. \cite{Kunihiro:1983ej,hatsuda1985soft}, this approximation is theoretically not self-consistent at the level of mean field approximation. \\
	\indent As shown in Refs. \cite{Kunihiro:1983ej,hatsuda1985soft}, to get a self-consistent result in the sense of mean field approximation, the contribution of the Fierz transformed Lagrangian (``exchange channel'') must be taken into account. Later, when the finite chemical potentials (in this situation, the vector-isoscalar channel interactions are very important) are involved, people usually put a vector-isoscalar channel interaction by hand to the original Lagrangian, so they must phenomenologically introduce a coupling parameter to reflect the strength of the explicit vector-isoscalar channel interaction \cite{buballa2005njl}. This will cause a serious problem, that is, the vector-isoscalar channel interaction is introduced phenomenologically, and the Lagrangian at this time is no longer the original Lagrangian. What is more serious is that the vector-isoscalar interaction introduced by hand is not theoretically self-consistent in the framework of the mean field approximation, because the Fierz transformed Lagrangian is not considered. \\
	\indent At the same time, if people want to discuss the problems related to the axialvector chemical potential (under these circumstances, the axialvector channel interaction plays an extremely important role), then people must artificially introduce the axialvector channel interaction. So, the introduction of a term by hand is quite arbitrary and will make the method lose its reliability \cite{yang2019qcd}. In order to overcome the arbitrariness brought by the above method, the self-consistent mean field approximation method is necessary, which can release all the interaction channels hidden in the original Lagrangian by the Fierz transformation \cite{su2020color}. Because the Fierz transformation is an equivalent transformation, the new Lagrangian $\Lagr_R=(1-\alpha)\Lagr+\alpha\Lagr_F$  used in this paper is equivalent to the original NJL model, which is another advantage of our method. \\
	\indent Actually, there are many studies that have considered the contribution of an explicit Fierz transformed term and have chosen the weight of vector-isoscalar channel factor $\alpha$ as 0.5 \cite{klevansky1992nambu,masayuki1989chiral}.  However, there is no physical basis supporting that the ``direct'' channel and the ``exchange'' channel have the same weight.
	In principle, the value of $\alpha$ could be constrained or hinted from related experiments and neutron star observations.
	On the other hand, in the commonly used NJL model, $\mu_0$ (when the chemical potential $\mu$ is smaller than the critical value $\mu_0$, the quark number density becomes zero. ) is very close to $\mu_c$ (the critical chemical potential of chiral restoration). It means that a vacuum phase transition occurs shortly after the baryon is excited from the vacuum, which is physically unreasonable \cite{zhao2019current,baym2018hadrons}. This long-standing problem can be well solved by adopting our self-consistent mean field approximation.

	\indent Based on the new Lagrangian of Eq. (\ref{E5}), the three-flavor quark gap equation is then given by
	\begin{align}
		\label{E6}M_i=&m_{i}-4G\sigma _i+2K\sigma _j\sigma _k ,\\
		\label{E7}\tilde{\mu_i}=&\mu_{i}-\frac{2}{3}\alpha G'(n_u+n_d+n_s),
	\end{align}
	where $G'=\frac{G}{1-\alpha}$, and $\tilde{\mu_i}$ is the effective chemical potential and $ M_i$ is the constituent quark mass. Similarly, $\mu_{i}$ is the chemical potential and $m_{i}$ is the quark mass.  Because  NJL model is  non-renormalizable, in this study, we employ  three momentum cutoff scheme to  regulate the divergence.  The quark condensation and number density in three momentum cutoff scheme is given by:
	\begin{align}
		\label{E8}\sigma_i=&-\frac{N_cM_i}{\pi^2}\int_0^\Lambda
		\frac{p^2}{E_{p,i}}[1-(e^{\frac{E_{p,i}-\tilde{\mu_i}}{T}}+1)^{-1}\\\notag&-(e^{\frac{E_{p,i}+\tilde{\mu_i}}{T}}+1)^{-1}] dp ,
	\end{align}
	\begin{align}
		\label{E9}n_i=\frac{N_c}{\pi^2}\int_0^\Lambda p^2[\frac{1}{e^{\frac{E_{p,i}-\tilde{\mu_i}}{T}}+1}-\frac{1}{e^{\frac{E_{p,i}+\tilde{\mu_i}}{T}}+1}] dp.
	\end{align}
	
	The parameters of NJL we adopted are shown in Table \ref{pa}, which are consistent with the experimental results of $\pi, \ K \ and\  \eta$ meson masses  and decay constants \cite{pereira2016two}.
	{
		\begin{table}[ht]
			\caption{Parameters adopted in our calculations}
			\begin{tabular}{c|c|c|c|c}
				\hline
				$m_u[MeV]$ &$m_s[MeV]$ &$\Lambda_{UV}[MeV]$ &$G\Lambda_{UV}^2$ &$K\Lambda_{UV}^5$  \\
				\hline
				5.5 &135.7 &630.1 &1.781 &9.29\\
				\hline
			\end{tabular}
			\label{pa}
	\end{table}}
	
	We can use the method of finite temperature field theory \cite{kapusta2006finite}  to write the grand canonical potential density as:
	\begin{align}\label{TP}
		\Omega=&\frac{-N_cT}{\pi^2}\sum_{i=u,d,s}\int p^2
		[\frac{E_{p,i}}{T}+ln(1+e^{-\frac{E_{p,i}+\tilde{\mu_i}}{T}})\\ \notag&+ln(1+e^{-\frac{E_{p,i}-\tilde{\mu_i}}{T}})]dp
		+2G(\sigma_u+\sigma_d+\sigma_s)\\ \notag&-\frac{\alpha G'}{3}(n_u+n_d+n_s)^2-4K\sigma_u\sigma_d\sigma_s+\Omega_0.
	\end{align}
	From this, the pressure $p=-\Omega$ and the energy density $\epsilon=-p+\sum_i\mu_{i}n_i$ can be easily obtained.
	In addition, the baryon number and electric charge are conserved for neutron stars \cite{glendenning2012compact}. Therefore, we need to include the following  equilibrium-condition:
	\[ \begin{cases} \frac{2}{3}n_u-\frac{1}{3}n_d-\frac{1}{3}n_s-n_e=0 ,\\ \mu_e=\mu_\mu=\mu_d-\mu_u , \\ \mu_d=\mu_s.\end{cases} \]\\

\begin{figure}[ht]
		\centering\includegraphics[width=1\linewidth]{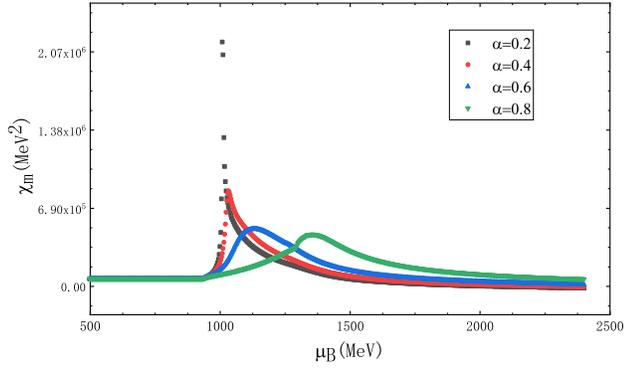}
		\caption{The chiral susceptibility as a function of baryon chemical potential at zero temperature, for different $\alpha$. }
		\label{chi1}
	\end{figure}

\begin{figure}[ht]
		\centering\includegraphics[width=1\linewidth]{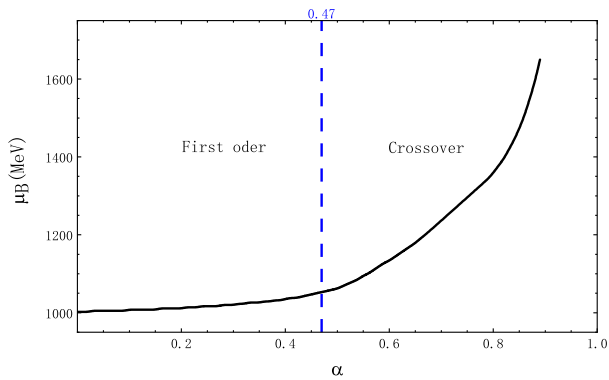}
		\caption{The critical (pseudo-critical) chemical potential as a function of $\alpha $ at zero temperature. When $\alpha < 0.47$, it is a first order phase transition. But when $\alpha > 0.47$, the phase transition is a crossover.}
		\label{chi2}
	\end{figure}

	\indent The NJL model with zero bare quark mass at high energy scale, keeps the chiral symmetry. With the decrease of energy, the quarks and anti-quarks condensate together to give quarks dynamical mass, thus the chiral symmetry is broken  at low energy scale.  To illustrate this phase transition more clearly, we can use the chiral
	susceptibility to reveal the order of phase transition and the corresponding physical condition\cite{AOKI200646,Fukushima:2010bq}. The chiral susceptibility is defined as :
	\begin{align}
		\chi_m=-\frac{\partial\sigma_u}{\partial m_{u}}.
	\end{align}
	\indent We show the results of chiral susceptibility for different $\alpha$ in Fig. \ref{chi1}.  It can be seen that when $\alpha$ is small, there is a singular point on the susceptibility curve. It corresponds to a first order phase transition occurring at a particular critical chemical potential. On the other hand, when $\alpha$ is large enough ($\alpha > 0.47$),  the curve becomes relatively smooth. It means that the phase transition changes to a crossover, and the chemical potential corresponding to the peak point is regarded as ``pseudo-critical chemical potential''.  Fig. \ref{chi1} also shows that  the critical (pseudo-critical) chemical potential gradually becomes larger with the increase of the parameter $\alpha$.  We can see this more clearly in Fig. \ref{chi2}.  The above results indicate that our quark model can be used to describe different kinds of phase transitions. By evaluating the parameter $\alpha$ properly, it can present a good approximation for various hadron models.  \\
	
\begin{figure}[ht]
		\centering\includegraphics[width=1\linewidth]{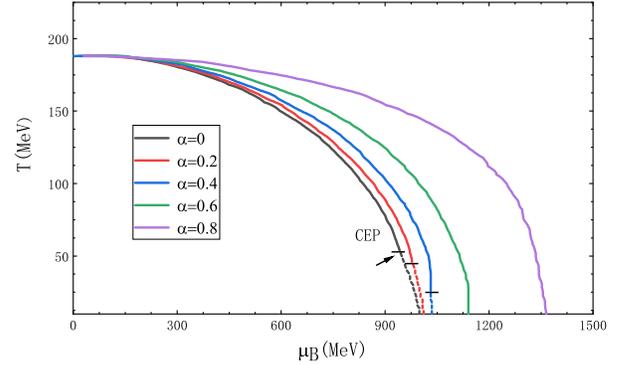}
		\caption{The QCD phase diagram on the $T-\mu$ plane for different $\alpha$ for the conditions of neutron stars. }
		\label{phase}
	\end{figure}

	\indent Similarly, using the chiral susceptibility we can plot the
	QCD phase diagram for different $\alpha$. The result is shown in
	Fig. \ref{phase}. In particular, we are interested in  the
	critical end point (CEP) which denotes a critical point connecting
	two different types of  transitions. It is one of the most
	important features of QCD phase diagram.  However,  whether there
	is a CEP in the phase diagram or not is still highly debated in
	the literature and it depends on the chosen physical environment \cite{Costa:2013zca}. Our model indicates that there is CEP for a small
	$\alpha$ but no CEP exists for a large $\alpha$.
	
\begin{figure}[ht]
		\centering\includegraphics[width=1\linewidth]{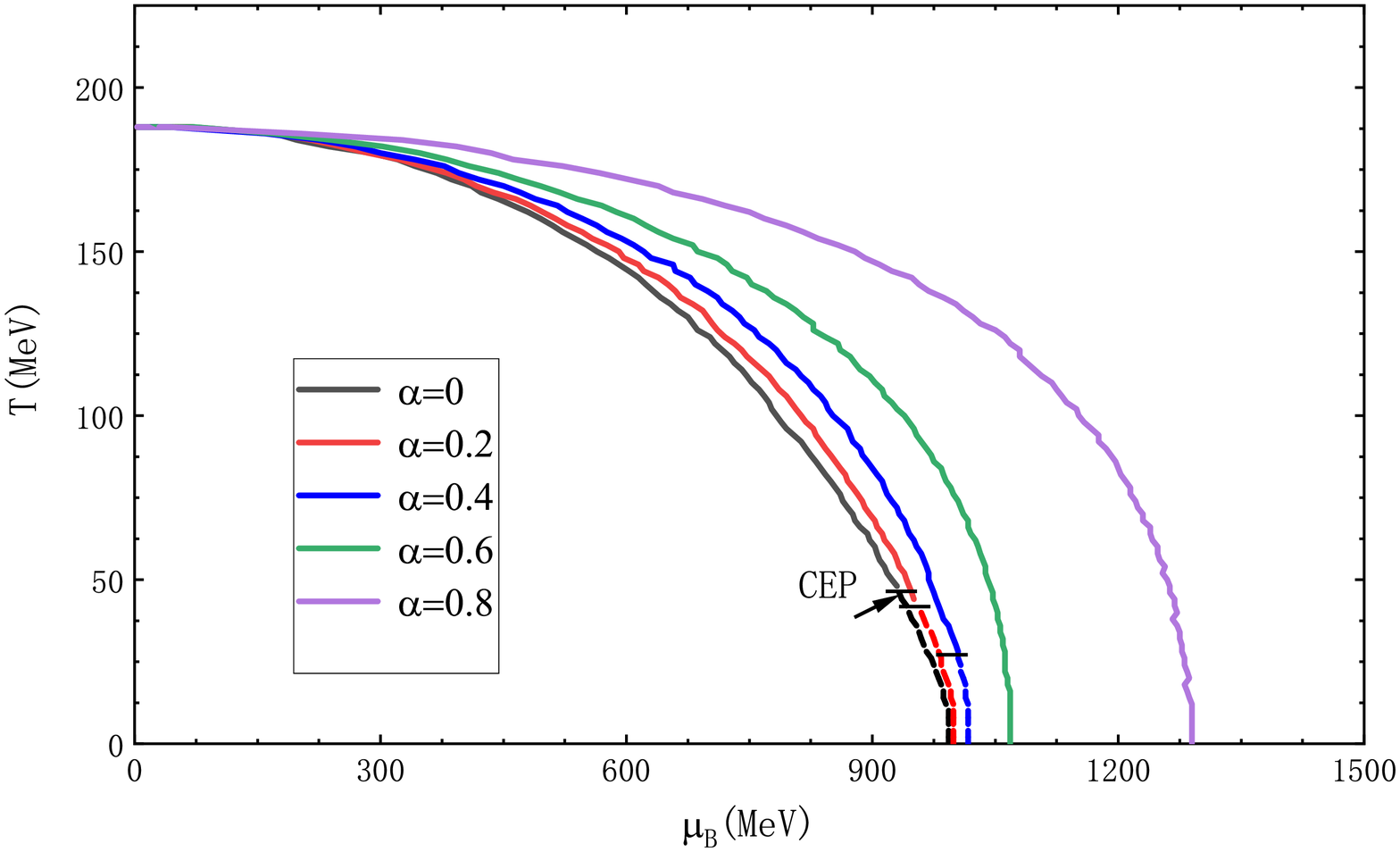}
		\caption{The QCD phase diagram on the $T-\mu$ plane for different $\alpha$,
				under the condition of $n_q=0.4n_B$ and $\beta$ equilibrium. When $\alpha>0.58$,
				there is no CEP in our model. }
		\label{phase04}
	\end{figure}

	More restrictions on $\alpha$ and the type of
		phase transition could be hinted by  the experiments of RHIC. Note
		that the above calculations about the QCD phase diagram are in the
		conditions of neutron stars. For the experiments of RHIC, the
		conditions are quite different. An important difference is that
		the ratio of electric charge over baryon is special and is
		approximately 0.4, i.e., $n_q = 0.4 n_B$. Second, the strangeness
		neutrality requires the density of strange quark ($n_s$) to be
		zero. Additionally, the experiments of RHIC are not in $\beta$
		equilibrium, but are connected with an ambiguous chemical
		potential condition. However, comparing with the special electric
		charge condition, such a chemical potential condition will have
		less influence on the QCD phase diagram. To compare with the RHIC
		experiments, we have calculated the QCD phase diagram under the
		above experimental conditions (but still with $\beta$
		equilibrium). The results are shown in Fig. \ref{phase04}. In
		these cases, our model predicts that there will be no CEP when
		$\alpha>0.58$ in the phase diagram. 
	
	\section{hybrid STARS WITH A CROSSOVER EOS}\label{s3}
\begin{figure}[ht]
		\centering\includegraphics[width=1\linewidth]{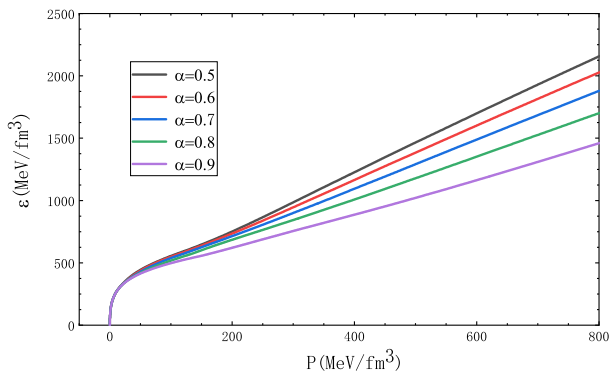}
		\caption{The crossover equation of state for different $\alpha$.}
		\label{pe}
	\end{figure}

	The EOS plays a critical role in calculating  the mass-radius relation and tidal deformability ($\Lambda$) of neutron stars.  Using Eq. (\ref{TP}), we can get the quark section of EOS. The hybrid star also includes a hadron section, but the NJL model is difficult to describe the hadron state at low densities. Thus we use the APR EOS for hadronic matter, which is a soft EOS \cite{Akmal:1998cf}. It has been adopted by many authros to describe hybrid stars at low densities\cite{Lau:2017qtz,Marczenko:2018jui}.  For the whole EOS, we use a simple mathematical method to link the hadron and quark section:
	\begin{align}
		&p=s(\mu)p_H(\mu)+(1-s(\mu))p_Q(\mu),\\
		&n=\frac{dp}{d\mu}=s(\mu)n_H(\mu)+(1-s(\mu))n_Q(\mu)+\frac{ds}{d\mu}(p_H-p_Q).
	\end{align}
	Here $p_H$ and $p_Q$ is the pressure of hadron and quark section, $p$ is the total pressure. The case is similar for the density $n$. The function $s$ is defined as:
	\begin{align}
		s(\mu)=e^{-\frac{\mu-\mu_0}{\mu_1}}.
	\end{align}
	Here the parameter $\mu_0=923$ MeV corresponds to the chemical potential that  nucleon matter  begins to appear in the ARP EOS model. The parameter $\mu_1=1200$ MeV we assume here can make  the contribution of hadron matter reduce and the contribution of quark matter increase smoothly when the density changes from $2n_0$ to $8n_0$, which means it is a crossover between quark and hadron matter. So, a reasonable requirement is that the EOS of quark matter also has a crossover, which means $\alpha>0.47$ here. The whole EOS is shown in Fig. \ref{pe}, from which it can be seen that a larger $\alpha$ will lead  to a harder EOS.
	
	\begin{figure}[ht]
		\centering\includegraphics[width=1\linewidth]{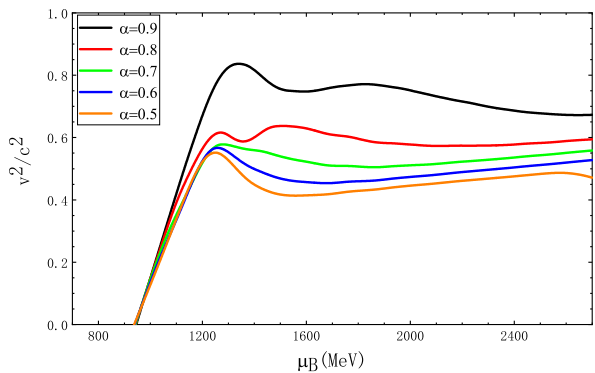}
		\caption{The square of sound velocity for different $\alpha$.}
		\label{sv}
	\end{figure}
	
\begin{figure}[ht]
		\centering\includegraphics[width=1\linewidth]{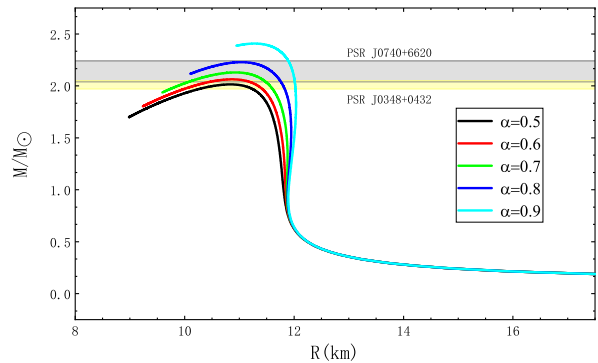}
		\caption{Mass-Radii relations of hybrid stars for different $\alpha$. The maximum mass increases with $\alpha $. Masses are in unites of $ M_{\odot}$.
		}
		\label{mr}
	\end{figure}

	\indent To study the stiffness of the system in detail, we calculate the square of sound velocity
	\begin{align}
		v^2=\frac{dp}{d\varepsilon},
	\end{align}
	which can reflect the stiffness of the system. The results are plot in Fig. 5. From  this figure, we can see that a larger $\alpha$ leads to a larger sound velocity, which demonstrates that the stiffness of the EOS will increase with $\alpha$. \\
	\indent  We then study  the mass-radius relation  by  solving the Tolman-Oppenheimer-Volkoff equations (in the natural unit system)
	\begin{align}
		&\frac{dp(r)}{dr}=-\frac{(\varepsilon+p)(M+4\pi r^3p)}{r(r-2M)} ,\\
		&\frac{dM(r)}{dr}=4\pi r^2\varepsilon.
	\end{align}
	In addition, we use the small crust approximation\cite{2016Neutron} to describe the crust of  our hybrid stars. To show the influence of  $\alpha$ on the neutron star, the mass-radius relation of hybrid stars for different values of $\alpha$ is presented in Fig. \ref{mr}.  Astronomical observations show that the masses of  PSR J0384+0432 and PSR J1614-2230 are $2.01\pm0.04 \ M_{\odot}$ and $1.928\pm0.017 \ M_{\odot}$ respectively. Our result indicates that when $\alpha=0.5$, the maximum {strange matter star} is $2.01 \ M_{\odot}$, which is consistent with observations. However, $\alpha=0.5$ is difficult to meet the requirement of PSR J0740+6620 $(2.14\pm0.10 \ M_{\odot})$.  Thus we may need an even larger $\alpha$, such as in the range of $0.6\sim0.8$. As shown in Fig. \ref{mr}, the maximum mass of hybrid stars increases with the parameter $\alpha$.  If a larger mass of neutron star is observed, then a larger $\alpha$ will be preferred.
	
\begin{figure}[ht]
		\centering\includegraphics[width=1\linewidth]{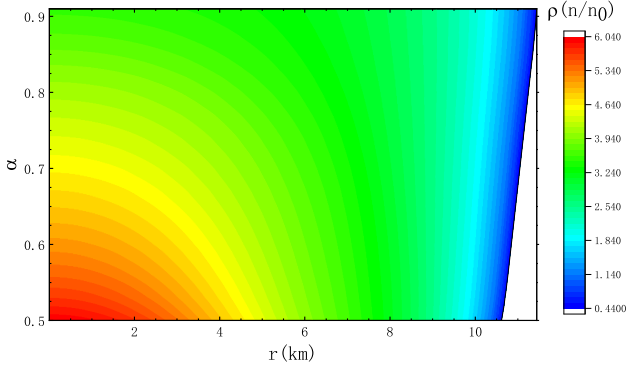}
		\caption{The density distribution in a $2M_\odot$ hybrid star. The color depth denotes the density value.
		}
		\label{rn}
	\end{figure}

	\indent Another astronomical phenomenon that can be engaged to limit the range of $\alpha$ is the cooling rate of neutron stars. The direct Urca  process provides the fastest neutrino emission in nucleon matter and quark matter. It is the main cooling mechanism in neutron stars \cite{Yakovlev:2004iq}. The neutrino emission rate is density-dependent. A high density will lead to a high emission rate, which makes the cooling of neutron stars faster. This could give us more restrictions on the EOS. We plot in Fig. \ref{rn} the density distribution inside a two solar mass hybrid star. We can see  that a small $\alpha$ results in a high density, which will lead to a quicker cooling rate. A more detailed comparison is beyond the scope of  this study and will be studied later.
	
	\begin{table}[ht]
		\caption{Properties of  hybrid stars in our framework}
		\centering
		\begin{tabular}{lllllp{10em}}
			\hline
			\hline
			$\alpha$ \ \ &$M_{max}(M_\odot)$ \ &R(km) &$R_{1. 4}(km)$ &$\Lambda(1. 4M_\odot)$ \\
			\hline
			0.5 \ \ &2.01 &10.81 &11.72  &312\\
			0.6 \ \ &2.06 &10.84 &11.80  &332\\
			0.7 \ \ &2.17 &10.92 &11.87  & 348\\
			0.8 \ \ &2.22 &11.03 &11.93  &363\\
			0.9 \ \ &2.40 &11.28 &11.98  &377\\
			\hline
			\hline
		\end{tabular}
		\label{T1}
	\end{table}

	During the inspiral and merger of two stars, the ratio of each star's induced mass quadrupole moment to the tidal field is defined as the tidal deformability $(\Lambda)$, and the Love number $k_2$ describes the distortion of the surface of a star. The relation between the tidal deformability $(\Lambda)$ and the $l=2$ dimensionless tidal Love number $k_2$ is  (in the natural unit system)
	\begin{align}
		k_2=\frac{3}{2}\Lambda(\frac{M}{R})^5.
	\end{align}
	According to Ref. \cite{damour2009relativistic}, $k_2$ can be calculated as
	\begin{align}
		k_2=&\frac{8C^5}{5}(1-2C^2)[2+2C(y-1)-y]\notag\\
		&\times\{2C[6-3y+3C(5y-8)]\notag\\
		&+4C^3[13-11y+C(3y-2)+2C^2(1+y)]\notag\\
		&+3(1-2C)^2[2-y+2C(y-1)]ln(1-2C)\}^{-1} ,
	\end{align}
	where $C=M/R$  is the compactness of the star, and
	\begin{align}
		y=\frac{R\beta(R)}{H(R)}-\frac{4\pi R^3 \varepsilon_0}{M}
	\end{align}
	is related to  the metric variable $H$ and surface energy density $\varepsilon_0$. The metric variable $H$ depends on the EOS and can be obtained by integrating two differential equations
	
	\begin{align}
		\frac{dH(r)}{dr}=\beta  ,
	\end{align}
	\begin{align}
		\label{E20}\frac{d\beta(r)}{dr}=&2(1-2\frac{M}{r})^{-1}H\{-2\pi[5\varepsilon+9p+f(\varepsilon+p)]\notag\\ &+\frac{3}{r^2}+2(1-2\frac{M}{r})^{-1}(\frac{M}{r^2}+4\pi r p)^2\}\notag\\ &+\frac{2\beta}{r}(1-2\frac{M}{r})^{-1}[-1+\frac{M}{r}+2\pi r^2 (\varepsilon-p)],
	\end{align}
	where $f$ is defined as
	\begin{align}
		\label{E21}f=\frac{d\varepsilon}{dp}.
	\end{align}
	We can integrate Eqs. (\ref{E20}) and (\ref{E21})  from the center via expansions $H(r)=a_0r^2$ and $\beta(r)=2a_0r$  for $r\ll R$ with constant $a_0$. Since $a_0$ can be reduced in the expression of the Love number $k_2$, here we take $a_0=1$, \\
	\indent We have calculated the main parameters of  hybrid stars for different $\alpha$. The results are shown in Table \ref{T1}. For a $1. 4M_\odot$  low-spin  star, the tidal deformability ($\Lambda$) has been constrained in a range of (200, 800) through gravitational wave observations \cite{doi:10.1063/1.5117803}. It can be seen that the results of our hybrid stars are consistent with the observational requirement.

	\section{ SUMMARY AND DISCUSSION }\label{s4}
	In this paper, we adopt a self-consistent mean  field
	approximation to discuss the three-flavor NJL model. A  free
	parameter $\alpha$ is introduced to denote the weight of different
	interaction channels. Using this method, we study the phase
	transition properties by  calculating  chiral susceptibility. It
	is found that the (pseudo) critical chemical potential increases
	with $\alpha$.  We plot the QCD phase diagram
		under the condition applicable for neutron stars as well as the
		condition for heavy ion collision experiments. It is found that
		there is no CEP when $\alpha>0.47$ for neutron stars. But for the
		heavy ion collision experiments,  there will be no CEP when $\alpha>0.58$. 
	We discuss how the stiffness of EOS changes with $\alpha$  by examining  the sound velocity.
	It is found that the phase transition of quark matter to hadron matter is a crossover when $\alpha>0.47$
	in neutron stars, thus we need a crossover connection method to describe the hybrid star. The corresponding EOS can also support a more massive star since it gets harder with a larger $\alpha $. The density-radius diagram of a $2M_\odot$ hybrid star is plot to show the effects of different parameters. Finally we calculate the tidal deformability and show that our EOS can meet the requirements of gravitational wave observations. \\
	\indent To be specific, the EOS with $\alpha=0.5$ yields a maximum compact star of $2.01 \ M_\odot$, which can  match the masses of PSR J1614-22300  $(1. 928\pm 0. 017 \ M_\odot)$ \cite{Fonseca:2016tux} and PSR J0348+0432 $(2. 01\pm0. 04 \ M_\odot)$\cite{Antoniadis:2013pzd}. However, $\alpha=0.5$ is difficult to satisfy PSR J0740+6620 $(2.14\pm0.10 \ M_{\odot})$\cite{NANOGrav:2019jur} and PSR J2215+5135 $(2.27^{+0.17}_{-0.15})$\cite{Linares:2018ppq}. A quark model with a larger $\alpha$ may provide a harder EOS to support a compact star with a larger mass. A more mild phase transition, which means a larger $\alpha$ in this paper, makes quarks have a larger energy at the same density due to interactions with each other.  As a result, $\alpha>0.7$ is preferred to satisfy PSR J0740+6620 $(2.14\pm0.10 \ M_{\odot})$ in our framework. More astronomical observations will help to further constrain the value of $\alpha$. We calculate the density-radius distribution because the density of matter may affect  the cooling rate of neutron stars. Measuring the temperature of neutron stars may also help to constrain $\alpha$. \\

	\section*{Acknowledgements}
	We thank the anonymous referee for helpful comments and
	suggestions. This work is supported by National SKA Program of China No. 2020SKA0120300, by the National Natural
	Science Foundation of China (Grant Nos. 11873030, 12041306,
	U1938201), by
	the National Key R\&D Program of China (2021YFA0718500), and by
	the science research grants from the China Manned Space Project
	with NO. CMS-CSST-2021-B11.
	\bibliographystyle{apsrev4-1}
	\bibliography{MYREF}

\begin{thebibliography}{57}%
\makeatletter
\providecommand \@ifxundefined [1]{%
 \@ifx{#1\undefined}
}%
\providecommand \@ifnum [1]{%
 \ifnum #1\expandafter \@firstoftwo
 \else \expandafter \@secondoftwo
 \fi
}%
\providecommand \@ifx [1]{%
 \ifx #1\expandafter \@firstoftwo
 \else \expandafter \@secondoftwo
 \fi
}%
\providecommand \natexlab [1]{#1}%
\providecommand \enquote  [1]{``#1''}%
\providecommand \bibnamefont  [1]{#1}%
\providecommand \bibfnamefont [1]{#1}%
\providecommand \citenamefont [1]{#1}%
\providecommand \href@noop [0]{\@secondoftwo}%
\providecommand \href [0]{\begingroup \@sanitize@url \@href}%
\providecommand \@href[1]{\@@startlink{#1}\@@href}%
\providecommand \@@href[1]{\endgroup#1\@@endlink}%
\providecommand \@sanitize@url [0]{\catcode `\\12\catcode `\$12\catcode
  `\&12\catcode `\#12\catcode `\^12\catcode `\_12\catcode `\%12\relax}%
\providecommand \@@startlink[1]{}%
\providecommand \@@endlink[0]{}%
\providecommand \url  [0]{\begingroup\@sanitize@url \@url }%
\providecommand \@url [1]{\endgroup\@href {#1}{\urlprefix }}%
\providecommand \urlprefix  [0]{URL }%
\providecommand \Eprint [0]{\href }%
\providecommand \doibase [0]{http://dx.doi.org/}%
\providecommand \selectlanguage [0]{\@gobble}%
\providecommand \bibinfo  [0]{\@secondoftwo}%
\providecommand \bibfield  [0]{\@secondoftwo}%
\providecommand \translation [1]{[#1]}%
\providecommand \BibitemOpen [0]{}%
\providecommand \bibitemStop [0]{}%
\providecommand \bibitemNoStop [0]{.\EOS\space}%
\providecommand \EOS [0]{\spacefactor3000\relax}%
\providecommand \BibitemShut  [1]{\csname bibitem#1\endcsname}%
\let\auto@bib@innerbib\@empty
%</preamble>
\bibitem [{\citenamefont {Abbott}\ \emph {et~al.}(2017)\citenamefont {Abbott},
  \citenamefont {Abbott}, \citenamefont {Abbott}, \citenamefont {Acernese},
  \citenamefont {Ackley}, \citenamefont {Adams}, \citenamefont {Adams},
  \citenamefont {Addesso}, \citenamefont {Adhikari}, \citenamefont {Adya} \emph
  {et~al.}}]{abbott2017gw170817}%
  \BibitemOpen
  \bibfield  {author} {\bibinfo {author} {\bibfnamefont {B.~P.}\ \bibnamefont
  {Abbott}}, \bibinfo {author} {\bibfnamefont {R.}~\bibnamefont {Abbott}},
  \bibinfo {author} {\bibfnamefont {T.}~\bibnamefont {Abbott}}, \bibinfo
  {author} {\bibfnamefont {F.}~\bibnamefont {Acernese}}, \bibinfo {author}
  {\bibfnamefont {K.}~\bibnamefont {Ackley}}, \bibinfo {author} {\bibfnamefont
  {C.}~\bibnamefont {Adams}}, \bibinfo {author} {\bibfnamefont
  {T.}~\bibnamefont {Adams}}, \bibinfo {author} {\bibfnamefont
  {P.}~\bibnamefont {Addesso}}, \bibinfo {author} {\bibfnamefont
  {R.}~\bibnamefont {Adhikari}}, \bibinfo {author} {\bibfnamefont
  {V.}~\bibnamefont {Adya}},  \emph {et~al.},\ }\href@noop {} {\bibfield
  {journal} {\bibinfo  {journal} {Physical Review Letters}\ }\textbf {\bibinfo
  {volume} {119}},\ \bibinfo {pages} {161101} (\bibinfo {year}
  {2017})}\BibitemShut {NoStop}%
\bibitem [{\citenamefont {Damour}\ \emph {et~al.}(1992)\citenamefont {Damour},
  \citenamefont {Soffel},\ and\ \citenamefont {Xu}}]{damour1992general}%
  \BibitemOpen
  \bibfield  {author} {\bibinfo {author} {\bibfnamefont {T.}~\bibnamefont
  {Damour}}, \bibinfo {author} {\bibfnamefont {M.}~\bibnamefont {Soffel}}, \
  and\ \bibinfo {author} {\bibfnamefont {C.}~\bibnamefont {Xu}},\ }\href@noop
  {} {\bibfield  {journal} {\bibinfo  {journal} {Physical Review D}\ }\textbf
  {\bibinfo {volume} {45}},\ \bibinfo {pages} {1017} (\bibinfo {year}
  {1992})}\BibitemShut {NoStop}%
\bibitem [{\citenamefont {Flanagan}\ and\ \citenamefont
  {Hinderer}(2008)}]{flanagan2008constraining}%
  \BibitemOpen
  \bibfield  {author} {\bibinfo {author} {\bibfnamefont {{\'E}.~{\'E}.}\
  \bibnamefont {Flanagan}}\ and\ \bibinfo {author} {\bibfnamefont
  {T.}~\bibnamefont {Hinderer}},\ }\href@noop {} {\bibfield  {journal}
  {\bibinfo  {journal} {Physical Review D}\ }\textbf {\bibinfo {volume} {77}},\
  \bibinfo {pages} {021502(R)} (\bibinfo {year} {2008})}\BibitemShut {NoStop}%
\bibitem [{\citenamefont {Hinderer}(2008)}]{hinderer2008tidal}%
  \BibitemOpen
  \bibfield  {author} {\bibinfo {author} {\bibfnamefont {T.}~\bibnamefont
  {Hinderer}},\ }\href@noop {} {\bibfield  {journal} {\bibinfo  {journal} {The
  Astrophysical Journal}\ }\textbf {\bibinfo {volume} {677}},\ \bibinfo {pages}
  {1216} (\bibinfo {year} {2008})}\BibitemShut {NoStop}%
\bibitem [{\citenamefont {Dunne}(2004)}]{Dunne:2004nc}%
  \BibitemOpen
  \bibfield  {author} {\bibinfo {author} {\bibfnamefont {G.~V.}\ \bibnamefont
  {Dunne}},\ }\enquote {\bibinfo {title} {{Heisenberg-Euler effective
  Lagrangians: Basics and extensions}},}\ in\ \href {\doibase
  10.1142/9789812775344_0014} {\emph {\bibinfo {booktitle} {{From fields to
  strings: Circumnavigating theoretical physics. Ian Kogan memorial collection
  (3 volume set)}}}},\ \bibinfo {editor} {edited by\ \bibinfo {editor}
  {\bibfnamefont {M.}~\bibnamefont {Shifman}}, \bibinfo {editor} {\bibfnamefont
  {A.}~\bibnamefont {Vainshtein}}, \ and\ \bibinfo {editor} {\bibfnamefont
  {J.}~\bibnamefont {Wheater}}}\ (\bibinfo {year} {2004})\ pp.\ \bibinfo
  {pages} {445--522},\ \Eprint {http://arxiv.org/abs/hep-th/0406216}
  {arXiv:hep-th/0406216} \BibitemShut {NoStop}%
\bibitem [{\citenamefont {Berger}(2014)}]{Berger:2013jza}%
  \BibitemOpen
  \bibfield  {author} {\bibinfo {author} {\bibfnamefont {E.}~\bibnamefont
  {Berger}},\ }\href {\doibase 10.1146/annurev-astro-081913-035926} {\bibfield
  {journal} {\bibinfo  {journal} {Ann. Rev. Astron. Astrophys.}\ }\textbf
  {\bibinfo {volume} {52}},\ \bibinfo {pages} {43} (\bibinfo {year} {2014})},\
  \Eprint {http://arxiv.org/abs/1311.2603} {arXiv:1311.2603 [astro-ph.HE]}
  \BibitemShut {NoStop}%
\bibitem [{\citenamefont {Miller}\ \emph {et~al.}(2019)\citenamefont {Miller}
  \emph {et~al.}}]{Miller:2019cac}%
  \BibitemOpen
  \bibfield  {author} {\bibinfo {author} {\bibfnamefont {M.~C.}\ \bibnamefont
  {Miller}} \emph {et~al.},\ }\href {\doibase 10.3847/2041-8213/ab50c5}
  {\bibfield  {journal} {\bibinfo  {journal} {Astrophys. J. Lett.}\ }\textbf
  {\bibinfo {volume} {887}},\ \bibinfo {pages} {L24} (\bibinfo {year}
  {2019})},\ \Eprint {http://arxiv.org/abs/1912.05705} {arXiv:1912.05705
  [astro-ph.HE]} \BibitemShut {NoStop}%
\bibitem [{\citenamefont {Geng}\ \emph {et~al.}(2021)\citenamefont {Geng},
  \citenamefont {Li},\ and\ \citenamefont {Huang}}]{Geng:2021apl}%
  \BibitemOpen
  \bibfield  {author} {\bibinfo {author} {\bibfnamefont {J.}~\bibnamefont
  {Geng}}, \bibinfo {author} {\bibfnamefont {B.}~\bibnamefont {Li}}, \ and\
  \bibinfo {author} {\bibfnamefont {Y.}~\bibnamefont {Huang}},\ }\href
  {\doibase https://doi.org/10.1016/j.xinn.2021.100152} {\bibfield  {journal}
  {\bibinfo  {journal} {The Innovation}\ }\textbf {\bibinfo {volume} {2}},\
  \bibinfo {pages} {100152} (\bibinfo {year} {2021})}\BibitemShut {NoStop}%
\bibitem [{\citenamefont {Bethke}(2007)}]{bethke2007experimental}%
  \BibitemOpen
  \bibfield  {author} {\bibinfo {author} {\bibfnamefont {S.}~\bibnamefont
  {Bethke}},\ }\href@noop {} {\bibfield  {journal} {\bibinfo  {journal}
  {Progress in Particle and Nuclear Physics}\ }\textbf {\bibinfo {volume}
  {58}},\ \bibinfo {pages} {351} (\bibinfo {year} {2007})}\BibitemShut
  {NoStop}%
\bibitem [{\citenamefont {Vogl}\ and\ \citenamefont
  {Weise}(1991)}]{vogl1991nambu}%
  \BibitemOpen
  \bibfield  {author} {\bibinfo {author} {\bibfnamefont {U.}~\bibnamefont
  {Vogl}}\ and\ \bibinfo {author} {\bibfnamefont {W.}~\bibnamefont {Weise}},\
  }\href@noop {} {\bibfield  {journal} {\bibinfo  {journal} {Progress in
  Particle and Nuclear Physics}\ }\textbf {\bibinfo {volume} {27}},\ \bibinfo
  {pages} {195} (\bibinfo {year} {1991})}\BibitemShut {NoStop}%
\bibitem [{\citenamefont {Klevansky}(1992)}]{klevansky1992nambu}%
  \BibitemOpen
  \bibfield  {author} {\bibinfo {author} {\bibfnamefont {S.}~\bibnamefont
  {Klevansky}},\ }\href@noop {} {\bibfield  {journal} {\bibinfo  {journal}
  {Reviews of Modern Physics}\ }\textbf {\bibinfo {volume} {64}},\ \bibinfo
  {pages} {649} (\bibinfo {year} {1992})}\BibitemShut {NoStop}%
\bibitem [{\citenamefont {Hatsuda}\ and\ \citenamefont
  {Kunihiro}(1994)}]{hatsuda1994qcd}%
  \BibitemOpen
  \bibfield  {author} {\bibinfo {author} {\bibfnamefont {T.}~\bibnamefont
  {Hatsuda}}\ and\ \bibinfo {author} {\bibfnamefont {T.}~\bibnamefont
  {Kunihiro}},\ }\href@noop {} {\bibfield  {journal} {\bibinfo  {journal}
  {Physics Reports}\ }\textbf {\bibinfo {volume} {247}},\ \bibinfo {pages}
  {221} (\bibinfo {year} {1994})}\BibitemShut {NoStop}%
\bibitem [{\citenamefont {Fan}\ \emph {et~al.}(2017)\citenamefont {Fan},
  \citenamefont {Luo},\ and\ \citenamefont {Zong}}]{fan2017mapping}%
  \BibitemOpen
  \bibfield  {author} {\bibinfo {author} {\bibfnamefont {W.}~\bibnamefont
  {Fan}}, \bibinfo {author} {\bibfnamefont {X.}~\bibnamefont {Luo}}, \ and\
  \bibinfo {author} {\bibfnamefont {H.-S.}\ \bibnamefont {Zong}},\ }\href@noop
  {} {\bibfield  {journal} {\bibinfo  {journal} {International Journal of
  Modern Physics A}\ }\textbf {\bibinfo {volume} {32}},\ \bibinfo {pages}
  {1750061} (\bibinfo {year} {2017})}\BibitemShut {NoStop}%
\bibitem [{\citenamefont {Li}\ \emph {et~al.}(2019)\citenamefont {Li},
  \citenamefont {Yin},\ and\ \citenamefont {Zong}}]{li2019new}%
  \BibitemOpen
  \bibfield  {author} {\bibinfo {author} {\bibfnamefont {C.-M.}\ \bibnamefont
  {Li}}, \bibinfo {author} {\bibfnamefont {P.-L.}\ \bibnamefont {Yin}}, \ and\
  \bibinfo {author} {\bibfnamefont {H.-S.}\ \bibnamefont {Zong}},\ }\href@noop
  {} {\bibfield  {journal} {\bibinfo  {journal} {Physical Review D}\ }\textbf
  {\bibinfo {volume} {99}},\ \bibinfo {pages} {076006} (\bibinfo {year}
  {2019})}\BibitemShut {NoStop}%
\bibitem [{\citenamefont {Philipsen}(2013)}]{Philipsen:2012nu}%
  \BibitemOpen
  \bibfield  {author} {\bibinfo {author} {\bibfnamefont {O.}~\bibnamefont
  {Philipsen}},\ }\href {\doibase 10.1016/j.ppnp.2012.09.003} {\bibfield
  {journal} {\bibinfo  {journal} {Prog. Part. Nucl. Phys.}\ }\textbf {\bibinfo
  {volume} {70}},\ \bibinfo {pages} {55} (\bibinfo {year} {2013})},\ \Eprint
  {http://arxiv.org/abs/1207.5999} {arXiv:1207.5999 [hep-lat]} \BibitemShut
  {NoStop}%
\bibitem [{\citenamefont {Fodor}\ and\ \citenamefont
  {Katz}(2009)}]{Fodor:2009ax}%
  \BibitemOpen
  \bibfield  {author} {\bibinfo {author} {\bibfnamefont {Z.}~\bibnamefont
  {Fodor}}\ and\ \bibinfo {author} {\bibfnamefont {S.~D.}\ \bibnamefont
  {Katz}},\ }\href@noop {} {\  (\bibinfo {year} {2009})},\ \Eprint
  {http://arxiv.org/abs/0908.3341} {arXiv:0908.3341 [hep-ph]} \BibitemShut
  {NoStop}%
\bibitem [{\citenamefont {Gupta}\ \emph {et~al.}(2008)\citenamefont {Gupta},
  \citenamefont {Huebner},\ and\ \citenamefont {Kaczmarek}}]{Gupta:2007ax}%
  \BibitemOpen
  \bibfield  {author} {\bibinfo {author} {\bibfnamefont {S.}~\bibnamefont
  {Gupta}}, \bibinfo {author} {\bibfnamefont {K.}~\bibnamefont {Huebner}}, \
  and\ \bibinfo {author} {\bibfnamefont {O.}~\bibnamefont {Kaczmarek}},\ }\href
  {\doibase 10.1103/PhysRevD.77.034503} {\bibfield  {journal} {\bibinfo
  {journal} {Phys. Rev. D}\ }\textbf {\bibinfo {volume} {77}},\ \bibinfo
  {pages} {034503} (\bibinfo {year} {2008})},\ \Eprint
  {http://arxiv.org/abs/0711.2251} {arXiv:0711.2251 [hep-lat]} \BibitemShut
  {NoStop}%
\bibitem [{\citenamefont {Borsanyi}\ \emph {et~al.}(2014)\citenamefont
  {Borsanyi}, \citenamefont {Fodor}, \citenamefont {Hoelbling}, \citenamefont
  {Katz}, \citenamefont {Krieg},\ and\ \citenamefont
  {Szabo}}]{Borsanyi:2013bia}%
  \BibitemOpen
  \bibfield  {author} {\bibinfo {author} {\bibfnamefont {S.}~\bibnamefont
  {Borsanyi}}, \bibinfo {author} {\bibfnamefont {Z.}~\bibnamefont {Fodor}},
  \bibinfo {author} {\bibfnamefont {C.}~\bibnamefont {Hoelbling}}, \bibinfo
  {author} {\bibfnamefont {S.~D.}\ \bibnamefont {Katz}}, \bibinfo {author}
  {\bibfnamefont {S.}~\bibnamefont {Krieg}}, \ and\ \bibinfo {author}
  {\bibfnamefont {K.~K.}\ \bibnamefont {Szabo}},\ }\href {\doibase
  10.1016/j.physletb.2014.01.007} {\bibfield  {journal} {\bibinfo  {journal}
  {Phys. Lett. B}\ }\textbf {\bibinfo {volume} {730}},\ \bibinfo {pages} {99}
  (\bibinfo {year} {2014})},\ \Eprint {http://arxiv.org/abs/1309.5258}
  {arXiv:1309.5258 [hep-lat]} \BibitemShut {NoStop}%
\bibitem [{\citenamefont {Aggarwal}\ \emph {et~al.}(2010)\citenamefont
  {Aggarwal}, \citenamefont {Ahammed}, \citenamefont {Alakhverdyants},
  \citenamefont {Alekseev}, \citenamefont {Alford}, \citenamefont {Anderson},
  \citenamefont {Arkhipkin}, \citenamefont {Averichev}, \citenamefont
  {Balewski}, \citenamefont {Barnby} \emph {et~al.}}]{aggarwal2010higher}%
  \BibitemOpen
  \bibfield  {author} {\bibinfo {author} {\bibfnamefont {M.}~\bibnamefont
  {Aggarwal}}, \bibinfo {author} {\bibfnamefont {Z.}~\bibnamefont {Ahammed}},
  \bibinfo {author} {\bibfnamefont {A.}~\bibnamefont {Alakhverdyants}},
  \bibinfo {author} {\bibfnamefont {I.}~\bibnamefont {Alekseev}}, \bibinfo
  {author} {\bibfnamefont {J.}~\bibnamefont {Alford}}, \bibinfo {author}
  {\bibfnamefont {B.}~\bibnamefont {Anderson}}, \bibinfo {author}
  {\bibfnamefont {D.}~\bibnamefont {Arkhipkin}}, \bibinfo {author}
  {\bibfnamefont {G.}~\bibnamefont {Averichev}}, \bibinfo {author}
  {\bibfnamefont {J.}~\bibnamefont {Balewski}}, \bibinfo {author}
  {\bibfnamefont {L.}~\bibnamefont {Barnby}},  \emph {et~al.},\ }\href@noop {}
  {\bibfield  {journal} {\bibinfo  {journal} {Physical Review Letters}\
  }\textbf {\bibinfo {volume} {105}},\ \bibinfo {pages} {022302} (\bibinfo
  {year} {2010})}\BibitemShut {NoStop}%
\bibitem [{\citenamefont {Adamczyk}\ \emph {et~al.}(2014)\citenamefont
  {Adamczyk}, \citenamefont {Adkins}, \citenamefont {Agakishiev}, \citenamefont
  {Aggarwal}, \citenamefont {Ahammed}, \citenamefont {Alekseev}, \citenamefont
  {Alford}, \citenamefont {Anson}, \citenamefont {Aparin}, \citenamefont
  {Arkhipkin} \emph {et~al.}}]{adamczyk2014energy}%
  \BibitemOpen
  \bibfield  {author} {\bibinfo {author} {\bibfnamefont {L.}~\bibnamefont
  {Adamczyk}}, \bibinfo {author} {\bibfnamefont {J.}~\bibnamefont {Adkins}},
  \bibinfo {author} {\bibfnamefont {G.}~\bibnamefont {Agakishiev}}, \bibinfo
  {author} {\bibfnamefont {M.}~\bibnamefont {Aggarwal}}, \bibinfo {author}
  {\bibfnamefont {Z.}~\bibnamefont {Ahammed}}, \bibinfo {author} {\bibfnamefont
  {I.}~\bibnamefont {Alekseev}}, \bibinfo {author} {\bibfnamefont
  {J.}~\bibnamefont {Alford}}, \bibinfo {author} {\bibfnamefont
  {C.}~\bibnamefont {Anson}}, \bibinfo {author} {\bibfnamefont
  {A.}~\bibnamefont {Aparin}}, \bibinfo {author} {\bibfnamefont
  {D.}~\bibnamefont {Arkhipkin}},  \emph {et~al.},\ }\href@noop {} {\bibfield
  {journal} {\bibinfo  {journal} {Physical Review Letters}\ }\textbf {\bibinfo
  {volume} {112}},\ \bibinfo {pages} {032302} (\bibinfo {year}
  {2014})}\BibitemShut {NoStop}%
\bibitem [{\citenamefont {Adamczyk}\ \emph {et~al.}(2018)\citenamefont
  {Adamczyk}, \citenamefont {Adams}, \citenamefont {Adkins}, \citenamefont
  {Agakishiev}, \citenamefont {Aggarwal}, \citenamefont {Ahammed},
  \citenamefont {Ajitanand}, \citenamefont {Alekseev}, \citenamefont
  {Anderson}, \citenamefont {Aoyama} \emph {et~al.}}]{adamczyk2018collision}%
  \BibitemOpen
  \bibfield  {author} {\bibinfo {author} {\bibfnamefont {L.}~\bibnamefont
  {Adamczyk}}, \bibinfo {author} {\bibfnamefont {J.}~\bibnamefont {Adams}},
  \bibinfo {author} {\bibfnamefont {J.~K.}\ \bibnamefont {Adkins}}, \bibinfo
  {author} {\bibfnamefont {G.}~\bibnamefont {Agakishiev}}, \bibinfo {author}
  {\bibfnamefont {M.}~\bibnamefont {Aggarwal}}, \bibinfo {author}
  {\bibfnamefont {Z.}~\bibnamefont {Ahammed}}, \bibinfo {author} {\bibfnamefont
  {N.}~\bibnamefont {Ajitanand}}, \bibinfo {author} {\bibfnamefont
  {I.}~\bibnamefont {Alekseev}}, \bibinfo {author} {\bibfnamefont
  {D.}~\bibnamefont {Anderson}}, \bibinfo {author} {\bibfnamefont
  {R.}~\bibnamefont {Aoyama}},  \emph {et~al.},\ }\href@noop {} {\bibfield
  {journal} {\bibinfo  {journal} {Physics Letters B}\ }\textbf {\bibinfo
  {volume} {785}},\ \bibinfo {pages} {551} (\bibinfo {year}
  {2018})}\BibitemShut {NoStop}%
\bibitem [{\citenamefont {Wang}\ \emph {et~al.}(2019)\citenamefont {Wang},
  \citenamefont {Cao},\ and\ \citenamefont {Zong}}]{wang2019novel}%
  \BibitemOpen
  \bibfield  {author} {\bibinfo {author} {\bibfnamefont {F.}~\bibnamefont
  {Wang}}, \bibinfo {author} {\bibfnamefont {Y.}~\bibnamefont {Cao}}, \ and\
  \bibinfo {author} {\bibfnamefont {H.}~\bibnamefont {Zong}},\ }\href@noop {}
  {\bibfield  {journal} {\bibinfo  {journal} {Chinese Physics C}\ }\textbf
  {\bibinfo {volume} {43}},\ \bibinfo {pages} {084102} (\bibinfo {year}
  {2019})}\BibitemShut {NoStop}%
\bibitem [{\citenamefont {Baym}\ and\ \citenamefont
  {Chin}(1976)}]{baym1976can}%
  \BibitemOpen
  \bibfield  {author} {\bibinfo {author} {\bibfnamefont {G.}~\bibnamefont
  {Baym}}\ and\ \bibinfo {author} {\bibfnamefont {S.}~\bibnamefont {Chin}},\
  }\href@noop {} {\bibfield  {journal} {\bibinfo  {journal} {Physics Letters
  B}\ }\textbf {\bibinfo {volume} {62}},\ \bibinfo {pages} {241} (\bibinfo
  {year} {1976})}\BibitemShut {NoStop}%
\bibitem [{\citenamefont {Baym}\ \emph {et~al.}(2018)\citenamefont {Baym},
  \citenamefont {Hatsuda}, \citenamefont {Kojo}, \citenamefont {Powell},
  \citenamefont {Song},\ and\ \citenamefont {Takatsuka}}]{baym2018hadrons}%
  \BibitemOpen
  \bibfield  {author} {\bibinfo {author} {\bibfnamefont {G.}~\bibnamefont
  {Baym}}, \bibinfo {author} {\bibfnamefont {T.}~\bibnamefont {Hatsuda}},
  \bibinfo {author} {\bibfnamefont {T.}~\bibnamefont {Kojo}}, \bibinfo {author}
  {\bibfnamefont {P.~D.}\ \bibnamefont {Powell}}, \bibinfo {author}
  {\bibfnamefont {Y.}~\bibnamefont {Song}}, \ and\ \bibinfo {author}
  {\bibfnamefont {T.}~\bibnamefont {Takatsuka}},\ }\href@noop {} {\bibfield
  {journal} {\bibinfo  {journal} {Reports on Progress in Physics}\ }\textbf
  {\bibinfo {volume} {81}},\ \bibinfo {pages} {056902} (\bibinfo {year}
  {2018})}\BibitemShut {NoStop}%
\bibitem [{\citenamefont {Li}\ \emph {et~al.}(2018{\natexlab{a}})\citenamefont
  {Li}, \citenamefont {Yan}, \citenamefont {Geng}, \citenamefont {Huang},\ and\
  \citenamefont {Zong}}]{li2018constraints}%
  \BibitemOpen
  \bibfield  {author} {\bibinfo {author} {\bibfnamefont {C.-M.}\ \bibnamefont
  {Li}}, \bibinfo {author} {\bibfnamefont {Y.}~\bibnamefont {Yan}}, \bibinfo
  {author} {\bibfnamefont {J.-J.}\ \bibnamefont {Geng}}, \bibinfo {author}
  {\bibfnamefont {Y.-F.}\ \bibnamefont {Huang}}, \ and\ \bibinfo {author}
  {\bibfnamefont {H.-S.}\ \bibnamefont {Zong}},\ }\href@noop {} {\bibfield
  {journal} {\bibinfo  {journal} {Physical Review D}\ }\textbf {\bibinfo
  {volume} {98}},\ \bibinfo {pages} {083013} (\bibinfo {year}
  {2018}{\natexlab{a}})}\BibitemShut {NoStop}%
\bibitem [{\citenamefont {Bai}\ and\ \citenamefont
  {Liu}(2019)}]{doi:10.1063/1.5117820}%
  \BibitemOpen
  \bibfield  {author} {\bibinfo {author} {\bibfnamefont {Z.}~\bibnamefont
  {Bai}}\ and\ \bibinfo {author} {\bibfnamefont {Y.-X.}\ \bibnamefont {Liu}},\
  }\href@noop {} {\bibfield  {journal} {\bibinfo  {journal} {AIP Conference
  Proceedings}\ }\textbf {\bibinfo {volume} {2127}},\ \bibinfo {pages} {020030}
  (\bibinfo {year} {2019})}\BibitemShut {NoStop}%
\bibitem [{\citenamefont {Masayuki}\ and\ \citenamefont
  {Koichi}(1989)}]{masayuki1989chiral}%
  \BibitemOpen
  \bibfield  {author} {\bibinfo {author} {\bibfnamefont {A.}~\bibnamefont
  {Masayuki}}\ and\ \bibinfo {author} {\bibfnamefont {Y.}~\bibnamefont
  {Koichi}},\ }\href@noop {} {\bibfield  {journal} {\bibinfo  {journal}
  {Nuclear Physics A}\ }\textbf {\bibinfo {volume} {504}},\ \bibinfo {pages}
  {668} (\bibinfo {year} {1989})}\BibitemShut {NoStop}%
\bibitem [{\citenamefont {Andersen}\ and\ \citenamefont
  {Strickland}(2002)}]{andersen2002equation}%
  \BibitemOpen
  \bibfield  {author} {\bibinfo {author} {\bibfnamefont {J.~O.}\ \bibnamefont
  {Andersen}}\ and\ \bibinfo {author} {\bibfnamefont {M.}~\bibnamefont
  {Strickland}},\ }\href@noop {} {\bibfield  {journal} {\bibinfo  {journal}
  {Physical Review D}\ }\textbf {\bibinfo {volume} {66}},\ \bibinfo {pages}
  {105001} (\bibinfo {year} {2002})}\BibitemShut {NoStop}%
\bibitem [{\citenamefont {Fortin}\ \emph {et~al.}(2016)\citenamefont {Fortin},
  \citenamefont {Provid\^encia}, \citenamefont {Raduta}, \citenamefont
  {Gulminelli}, \citenamefont {Zdunik}, \citenamefont {Haensel},\ and\
  \citenamefont {Bejger}}]{fortin2016neutron}%
  \BibitemOpen
  \bibfield  {author} {\bibinfo {author} {\bibfnamefont {M.}~\bibnamefont
  {Fortin}}, \bibinfo {author} {\bibfnamefont {C.}~\bibnamefont
  {Provid\^encia}}, \bibinfo {author} {\bibfnamefont {A.~R.}\ \bibnamefont
  {Raduta}}, \bibinfo {author} {\bibfnamefont {F.}~\bibnamefont {Gulminelli}},
  \bibinfo {author} {\bibfnamefont {J.~L.}\ \bibnamefont {Zdunik}}, \bibinfo
  {author} {\bibfnamefont {P.}~\bibnamefont {Haensel}}, \ and\ \bibinfo
  {author} {\bibfnamefont {M.}~\bibnamefont {Bejger}},\ }\href {\doibase
  10.1103/PhysRevC.94.035804} {\bibfield  {journal} {\bibinfo  {journal} {Phys.
  Rev. C}\ }\textbf {\bibinfo {volume} {94}},\ \bibinfo {pages} {035804}
  (\bibinfo {year} {2016})}\BibitemShut {NoStop}%
\bibitem [{\citenamefont {Li}\ \emph {et~al.}(2018{\natexlab{b}})\citenamefont
  {Li}, \citenamefont {Zhang}, \citenamefont {Yan}, \citenamefont {Huang},\
  and\ \citenamefont {Zong}}]{li2018studies}%
  \BibitemOpen
  \bibfield  {author} {\bibinfo {author} {\bibfnamefont {C.-M.}\ \bibnamefont
  {Li}}, \bibinfo {author} {\bibfnamefont {J.-L.}\ \bibnamefont {Zhang}},
  \bibinfo {author} {\bibfnamefont {Y.}~\bibnamefont {Yan}}, \bibinfo {author}
  {\bibfnamefont {Y.-F.}\ \bibnamefont {Huang}}, \ and\ \bibinfo {author}
  {\bibfnamefont {H.-S.}\ \bibnamefont {Zong}},\ }\href@noop {} {\bibfield
  {journal} {\bibinfo  {journal} {Physical Review D}\ }\textbf {\bibinfo
  {volume} {97}},\ \bibinfo {pages} {103013} (\bibinfo {year}
  {2018}{\natexlab{b}})}\BibitemShut {NoStop}%
\bibitem [{\citenamefont {Chen}\ \emph {et~al.}(2011)\citenamefont {Chen},
  \citenamefont {Baldo}, \citenamefont {Burgio},\ and\ \citenamefont
  {Schulze}}]{Chen:2011my}%
  \BibitemOpen
  \bibfield  {author} {\bibinfo {author} {\bibfnamefont {H.}~\bibnamefont
  {Chen}}, \bibinfo {author} {\bibfnamefont {M.}~\bibnamefont {Baldo}},
  \bibinfo {author} {\bibfnamefont {G.~F.}\ \bibnamefont {Burgio}}, \ and\
  \bibinfo {author} {\bibfnamefont {H.~J.}\ \bibnamefont {Schulze}},\ }\href
  {\doibase 10.1103/PhysRevD.84.105023} {\bibfield  {journal} {\bibinfo
  {journal} {Phys. Rev. D}\ }\textbf {\bibinfo {volume} {84}},\ \bibinfo
  {pages} {105023} (\bibinfo {year} {2011})},\ \Eprint
  {http://arxiv.org/abs/1107.2497} {arXiv:1107.2497 [nucl-th]} \BibitemShut
  {NoStop}%
\bibitem [{\citenamefont {Masuda}\ \emph {et~al.}(2013)\citenamefont {Masuda},
  \citenamefont {Hatsuda},\ and\ \citenamefont {Takatsuka}}]{Masuda:2012ed}%
  \BibitemOpen
  \bibfield  {author} {\bibinfo {author} {\bibfnamefont {K.}~\bibnamefont
  {Masuda}}, \bibinfo {author} {\bibfnamefont {T.}~\bibnamefont {Hatsuda}}, \
  and\ \bibinfo {author} {\bibfnamefont {T.}~\bibnamefont {Takatsuka}},\ }\href
  {\doibase 10.1093/ptep/ptt045} {\bibfield  {journal} {\bibinfo  {journal}
  {PTEP}\ }\textbf {\bibinfo {volume} {2013}},\ \bibinfo {pages} {073D01}
  (\bibinfo {year} {2013})},\ \Eprint {http://arxiv.org/abs/1212.6803}
  {arXiv:1212.6803 [nucl-th]} \BibitemShut {NoStop}%
\bibitem [{\citenamefont {Kapusta}\ and\ \citenamefont
  {Welle}(2021)}]{PhysRevC.104.L012801}%
  \BibitemOpen
  \bibfield  {author} {\bibinfo {author} {\bibfnamefont {J.~I.}\ \bibnamefont
  {Kapusta}}\ and\ \bibinfo {author} {\bibfnamefont {T.}~\bibnamefont
  {Welle}},\ }\href {\doibase 10.1103/PhysRevC.104.L012801} {\bibfield
  {journal} {\bibinfo  {journal} {Phys. Rev. C}\ }\textbf {\bibinfo {volume}
  {104}},\ \bibinfo {pages} {L012801} (\bibinfo {year} {2021})}\BibitemShut
  {NoStop}%
\bibitem [{\citenamefont {Akmal}\ \emph {et~al.}(1998)\citenamefont {Akmal},
  \citenamefont {Pandharipande},\ and\ \citenamefont
  {Ravenhall}}]{Akmal:1998cf}%
  \BibitemOpen
  \bibfield  {author} {\bibinfo {author} {\bibfnamefont {A.}~\bibnamefont
  {Akmal}}, \bibinfo {author} {\bibfnamefont {V.~R.}\ \bibnamefont
  {Pandharipande}}, \ and\ \bibinfo {author} {\bibfnamefont {D.~G.}\
  \bibnamefont {Ravenhall}},\ }\href {\doibase 10.1103/PhysRevC.58.1804}
  {\bibfield  {journal} {\bibinfo  {journal} {Phys. Rev. C}\ }\textbf {\bibinfo
  {volume} {58}},\ \bibinfo {pages} {1804} (\bibinfo {year} {1998})},\ \Eprint
  {http://arxiv.org/abs/nucl-th/9804027} {arXiv:nucl-th/9804027} \BibitemShut
  {NoStop}%
\bibitem [{\citenamefont {Hatsuda}\ and\ \citenamefont
  {Kunihiro}(1985)}]{hatsuda1985soft}%
  \BibitemOpen
  \bibfield  {author} {\bibinfo {author} {\bibfnamefont {T.}~\bibnamefont
  {Hatsuda}}\ and\ \bibinfo {author} {\bibfnamefont {T.}~\bibnamefont
  {Kunihiro}},\ }\href@noop {} {\bibfield  {journal} {\bibinfo  {journal}
  {Progress of Theoretical Physics}\ }\textbf {\bibinfo {volume} {74}},\
  \bibinfo {pages} {765} (\bibinfo {year} {1985})}\BibitemShut {NoStop}%
\bibitem [{\citenamefont {Nambu}\ and\ \citenamefont
  {Jona-Lasinio}(1961)}]{Nambu:1961tp}%
  \BibitemOpen
  \bibfield  {author} {\bibinfo {author} {\bibfnamefont {Y.}~\bibnamefont
  {Nambu}}\ and\ \bibinfo {author} {\bibfnamefont {G.}~\bibnamefont
  {Jona-Lasinio}},\ }\href {\doibase 10.1103/PhysRev.122.345} {\bibfield
  {journal} {\bibinfo  {journal} {Phys. Rev.}\ }\textbf {\bibinfo {volume}
  {122}},\ \bibinfo {pages} {345} (\bibinfo {year} {1961})}\BibitemShut
  {NoStop}%
\bibitem [{\citenamefont {Kunihiro}\ and\ \citenamefont
  {Hatsuda}(1984)}]{Kunihiro:1983ej}%
  \BibitemOpen
  \bibfield  {author} {\bibinfo {author} {\bibfnamefont {T.}~\bibnamefont
  {Kunihiro}}\ and\ \bibinfo {author} {\bibfnamefont {T.}~\bibnamefont
  {Hatsuda}},\ }\href {\doibase 10.1143/PTP.71.1332} {\bibfield  {journal}
  {\bibinfo  {journal} {Prog. Theor. Phys.}\ }\textbf {\bibinfo {volume}
  {71}},\ \bibinfo {pages} {1332} (\bibinfo {year} {1984})}\BibitemShut
  {NoStop}%
\bibitem [{\citenamefont {Buballa}(2005)}]{buballa2005njl}%
  \BibitemOpen
  \bibfield  {author} {\bibinfo {author} {\bibfnamefont {M.}~\bibnamefont
  {Buballa}},\ }\href@noop {} {\bibfield  {journal} {\bibinfo  {journal}
  {Physics Reports}\ }\textbf {\bibinfo {volume} {407}},\ \bibinfo {pages}
  {205} (\bibinfo {year} {2005})}\BibitemShut {NoStop}%
\bibitem [{\citenamefont {Yang}\ \emph {et~al.}(2019)\citenamefont {Yang},
  \citenamefont {Luo},\ and\ \citenamefont {Zong}}]{yang2019qcd}%
  \BibitemOpen
  \bibfield  {author} {\bibinfo {author} {\bibfnamefont {L.-K.}\ \bibnamefont
  {Yang}}, \bibinfo {author} {\bibfnamefont {X.}~\bibnamefont {Luo}}, \ and\
  \bibinfo {author} {\bibfnamefont {H.-S.}\ \bibnamefont {Zong}},\ }\href
  {\doibase 10.1103/PhysRevD.100.094012} {\bibfield  {journal} {\bibinfo
  {journal} {Phys. Rev. D}\ }\textbf {\bibinfo {volume} {100}},\ \bibinfo
  {pages} {094012} (\bibinfo {year} {2019})},\ \Eprint
  {http://arxiv.org/abs/1910.13185} {arXiv:1910.13185 [nucl-th]} \BibitemShut
  {NoStop}%
\bibitem [{\citenamefont {Su}\ \emph {et~al.}(2020)\citenamefont {Su},
  \citenamefont {Shi}, \citenamefont {xia},\ and\ \citenamefont
  {Zong}}]{su2020color}%
  \BibitemOpen
  \bibfield  {author} {\bibinfo {author} {\bibfnamefont {L.-Q.}\ \bibnamefont
  {Su}}, \bibinfo {author} {\bibfnamefont {C.}~\bibnamefont {Shi}}, \bibinfo
  {author} {\bibfnamefont {Y.-H.}\ \bibnamefont {xia}}, \ and\ \bibinfo
  {author} {\bibfnamefont {H.}~\bibnamefont {Zong}},\ }\href@noop {} {\enquote
  {\bibinfo {title} {Color superconductivity with self-consistent njl-type
  model},}\ } (\bibinfo {year} {2020}),\ \Eprint
  {http://arxiv.org/abs/2008.07678} {arXiv:2008.07678 [nucl-th]} \BibitemShut
  {NoStop}%
\bibitem [{\citenamefont {Zhao}\ \emph {et~al.}(2019)\citenamefont {Zhao},
  \citenamefont {Zheng}, \citenamefont {Wang}, \citenamefont {Li},
  \citenamefont {Yan}, \citenamefont {Huang},\ and\ \citenamefont
  {Zong}}]{zhao2019current}%
  \BibitemOpen
  \bibfield  {author} {\bibinfo {author} {\bibfnamefont {T.}~\bibnamefont
  {Zhao}}, \bibinfo {author} {\bibfnamefont {W.}~\bibnamefont {Zheng}},
  \bibinfo {author} {\bibfnamefont {F.}~\bibnamefont {Wang}}, \bibinfo {author}
  {\bibfnamefont {C.-M.}\ \bibnamefont {Li}}, \bibinfo {author} {\bibfnamefont
  {Y.}~\bibnamefont {Yan}}, \bibinfo {author} {\bibfnamefont {Y.-F.}\
  \bibnamefont {Huang}}, \ and\ \bibinfo {author} {\bibfnamefont {H.-S.}\
  \bibnamefont {Zong}},\ }\href@noop {} {\bibfield  {journal} {\bibinfo
  {journal} {Physical Review D}\ }\textbf {\bibinfo {volume} {100}},\ \bibinfo
  {pages} {043018} (\bibinfo {year} {2019})}\BibitemShut {NoStop}%
\bibitem [{\citenamefont {Pereira}\ \emph {et~al.}(2016)\citenamefont
  {Pereira}, \citenamefont {Costa},\ and\ \citenamefont
  {Provid{\^e}ncia}}]{pereira2016two}%
  \BibitemOpen
  \bibfield  {author} {\bibinfo {author} {\bibfnamefont {R.~C.}\ \bibnamefont
  {Pereira}}, \bibinfo {author} {\bibfnamefont {P.}~\bibnamefont {Costa}}, \
  and\ \bibinfo {author} {\bibfnamefont {C.}~\bibnamefont {Provid{\^e}ncia}},\
  }\href@noop {} {\bibfield  {journal} {\bibinfo  {journal} {Physical Review
  D}\ }\textbf {\bibinfo {volume} {94}},\ \bibinfo {pages} {094001} (\bibinfo
  {year} {2016})}\BibitemShut {NoStop}%
\bibitem [{\citenamefont {Kapusta}\ and\ \citenamefont
  {Gale}(2006)}]{kapusta2006finite}%
  \BibitemOpen
  \bibfield  {author} {\bibinfo {author} {\bibfnamefont {J.}~\bibnamefont
  {Kapusta}}\ and\ \bibinfo {author} {\bibfnamefont {C.}~\bibnamefont {Gale}},\
  }\href {https://books.google.pt/books?id=rll8dJ2iTpsC} {\emph {\bibinfo
  {title} {Finite-Temperature Field Theory: Principles and Applications}}},\
  Cambridge Monographs on Mathematical Physics\ (\bibinfo  {publisher}
  {Cambridge University Press},\ \bibinfo {year} {2006})\BibitemShut {NoStop}%
\bibitem [{\citenamefont {Glendenning}(2012)}]{glendenning2012compact}%
  \BibitemOpen
  \bibfield  {author} {\bibinfo {author} {\bibfnamefont {N.~K.}\ \bibnamefont
  {Glendenning}},\ }\href@noop {} {\emph {\bibinfo {title} {Compact stars:
  Nuclear physics, particle physics and general relativity}}}\ (\bibinfo
  {publisher} {Springer Science \& Business Media},\ \bibinfo {year}
  {2012})\BibitemShut {NoStop}%
\bibitem [{\citenamefont {Aoki}\ \emph {et~al.}(2006)\citenamefont {Aoki},
  \citenamefont {Fodor}, \citenamefont {Katz},\ and\ \citenamefont
  {Szabó}}]{AOKI200646}%
  \BibitemOpen
  \bibfield  {author} {\bibinfo {author} {\bibfnamefont {Y.}~\bibnamefont
  {Aoki}}, \bibinfo {author} {\bibfnamefont {Z.}~\bibnamefont {Fodor}},
  \bibinfo {author} {\bibfnamefont {S.}~\bibnamefont {Katz}}, \ and\ \bibinfo
  {author} {\bibfnamefont {K.}~\bibnamefont {Szabó}},\ }\href {\doibase
  https://doi.org/10.1016/j.physletb.2006.10.021} {\bibfield  {journal}
  {\bibinfo  {journal} {Physics Letters B}\ }\textbf {\bibinfo {volume}
  {643}},\ \bibinfo {pages} {46} (\bibinfo {year} {2006})}\BibitemShut
  {NoStop}%
\bibitem [{\citenamefont {Fukushima}\ and\ \citenamefont
  {Hatsuda}(2011)}]{Fukushima:2010bq}%
  \BibitemOpen
  \bibfield  {author} {\bibinfo {author} {\bibfnamefont {K.}~\bibnamefont
  {Fukushima}}\ and\ \bibinfo {author} {\bibfnamefont {T.}~\bibnamefont
  {Hatsuda}},\ }\href {\doibase 10.1088/0034-4885/74/1/014001} {\bibfield
  {journal} {\bibinfo  {journal} {Rept. Prog. Phys.}\ }\textbf {\bibinfo
  {volume} {74}},\ \bibinfo {pages} {014001} (\bibinfo {year} {2011})},\
  \Eprint {http://arxiv.org/abs/1005.4814} {arXiv:1005.4814 [hep-ph]}
  \BibitemShut {NoStop}%
\bibitem [{\citenamefont {Costa}\ \emph {et~al.}(2014)\citenamefont {Costa},
  \citenamefont {Ferreira}, \citenamefont {Hansen}, \citenamefont {Menezes},\
  and\ \citenamefont {Provid\^encia}}]{Costa:2013zca}%
  \BibitemOpen
  \bibfield  {author} {\bibinfo {author} {\bibfnamefont {P.}~\bibnamefont
  {Costa}}, \bibinfo {author} {\bibfnamefont {M.}~\bibnamefont {Ferreira}},
  \bibinfo {author} {\bibfnamefont {H.}~\bibnamefont {Hansen}}, \bibinfo
  {author} {\bibfnamefont {D.~P.}\ \bibnamefont {Menezes}}, \ and\ \bibinfo
  {author} {\bibfnamefont {C.}~\bibnamefont {Provid\^encia}},\ }\href {\doibase
  10.1103/PhysRevD.89.056013} {\bibfield  {journal} {\bibinfo  {journal} {Phys.
  Rev. D}\ }\textbf {\bibinfo {volume} {89}},\ \bibinfo {pages} {056013}
  (\bibinfo {year} {2014})},\ \Eprint {http://arxiv.org/abs/1307.7894}
  {arXiv:1307.7894 [hep-ph]} \BibitemShut {NoStop}%
\bibitem [{\citenamefont {Lau}\ \emph {et~al.}(2017)\citenamefont {Lau},
  \citenamefont {Leung},\ and\ \citenamefont {Lin}}]{Lau:2017qtz}%
  \BibitemOpen
  \bibfield  {author} {\bibinfo {author} {\bibfnamefont {S.~Y.}\ \bibnamefont
  {Lau}}, \bibinfo {author} {\bibfnamefont {P.~T.}\ \bibnamefont {Leung}}, \
  and\ \bibinfo {author} {\bibfnamefont {L.~M.}\ \bibnamefont {Lin}},\ }\href
  {\doibase 10.1103/PhysRevD.95.101302} {\bibfield  {journal} {\bibinfo
  {journal} {Phys. Rev. D}\ }\textbf {\bibinfo {volume} {95}},\ \bibinfo
  {pages} {101302} (\bibinfo {year} {2017})},\ \Eprint
  {http://arxiv.org/abs/1705.01710} {arXiv:1705.01710 [astro-ph.HE]}
  \BibitemShut {NoStop}%
\bibitem [{\citenamefont {Marczenko}\ \emph {et~al.}(2018)\citenamefont
  {Marczenko}, \citenamefont {Blaschke}, \citenamefont {Redlich},\ and\
  \citenamefont {Sasaki}}]{Marczenko:2018jui}%
  \BibitemOpen
  \bibfield  {author} {\bibinfo {author} {\bibfnamefont {M.}~\bibnamefont
  {Marczenko}}, \bibinfo {author} {\bibfnamefont {D.}~\bibnamefont {Blaschke}},
  \bibinfo {author} {\bibfnamefont {K.}~\bibnamefont {Redlich}}, \ and\
  \bibinfo {author} {\bibfnamefont {C.}~\bibnamefont {Sasaki}},\ }\href
  {\doibase 10.1103/PhysRevD.98.103021} {\bibfield  {journal} {\bibinfo
  {journal} {Phys. Rev. D}\ }\textbf {\bibinfo {volume} {98}},\ \bibinfo
  {pages} {103021} (\bibinfo {year} {2018})},\ \Eprint
  {http://arxiv.org/abs/1805.06886} {arXiv:1805.06886 [nucl-th]} \BibitemShut
  {NoStop}%
\bibitem [{\citenamefont {Zdunik}\ \emph {et~al.}(2016)\citenamefont {Zdunik},
  \citenamefont {Fortin},\ and\ \citenamefont {Haensel}}]{2016Neutron}%
  \BibitemOpen
  \bibfield  {author} {\bibinfo {author} {\bibfnamefont {J.~L.}\ \bibnamefont
  {Zdunik}}, \bibinfo {author} {\bibfnamefont {M.}~\bibnamefont {Fortin}}, \
  and\ \bibinfo {author} {\bibfnamefont {P.}~\bibnamefont {Haensel}},\
  }\href@noop {} {\bibfield  {journal} {\bibinfo  {journal} {Astronomy and
  Astrophysics}\ }\textbf {\bibinfo {volume} {599}},\ \bibinfo {pages} {A119}
  (\bibinfo {year} {2016})}\BibitemShut {NoStop}%
\bibitem [{\citenamefont {Yakovlev}\ and\ \citenamefont
  {Pethick}(2004)}]{Yakovlev:2004iq}%
  \BibitemOpen
  \bibfield  {author} {\bibinfo {author} {\bibfnamefont {D.~G.}\ \bibnamefont
  {Yakovlev}}\ and\ \bibinfo {author} {\bibfnamefont {C.~J.}\ \bibnamefont
  {Pethick}},\ }\href {\doibase 10.1146/annurev.astro.42.053102.134013}
  {\bibfield  {journal} {\bibinfo  {journal} {Ann. Rev. Astron. Astrophys.}\
  }\textbf {\bibinfo {volume} {42}},\ \bibinfo {pages} {169} (\bibinfo {year}
  {2004})},\ \Eprint {http://arxiv.org/abs/astro-ph/0402143}
  {arXiv:astro-ph/0402143} \BibitemShut {NoStop}%
\bibitem [{\citenamefont {Damour}\ and\ \citenamefont
  {Nagar}(2009)}]{damour2009relativistic}%
  \BibitemOpen
  \bibfield  {author} {\bibinfo {author} {\bibfnamefont {T.}~\bibnamefont
  {Damour}}\ and\ \bibinfo {author} {\bibfnamefont {A.}~\bibnamefont {Nagar}},\
  }\href@noop {} {\bibfield  {journal} {\bibinfo  {journal} {Physical Review
  D}\ }\textbf {\bibinfo {volume} {80}},\ \bibinfo {pages} {084035} (\bibinfo
  {year} {2009})}\BibitemShut {NoStop}%
\bibitem [{\citenamefont {Bauswein}\ \emph {et~al.}(2019)\citenamefont
  {Bauswein}, \citenamefont {Bastian}, \citenamefont {Blaschke}, \citenamefont
  {Chatziioannou}, \citenamefont {Clark}, \citenamefont {Fischer},
  \citenamefont {Janka}, \citenamefont {Just}, \citenamefont {Oertel},\ and\
  \citenamefont {Stergioulas}}]{doi:10.1063/1.5117803}%
  \BibitemOpen
  \bibfield  {author} {\bibinfo {author} {\bibfnamefont {A.}~\bibnamefont
  {Bauswein}}, \bibinfo {author} {\bibfnamefont {N.-U.~F.}\ \bibnamefont
  {Bastian}}, \bibinfo {author} {\bibfnamefont {D.}~\bibnamefont {Blaschke}},
  \bibinfo {author} {\bibfnamefont {K.}~\bibnamefont {Chatziioannou}}, \bibinfo
  {author} {\bibfnamefont {J.~A.}\ \bibnamefont {Clark}}, \bibinfo {author}
  {\bibfnamefont {T.}~\bibnamefont {Fischer}}, \bibinfo {author} {\bibfnamefont
  {H.-T.}\ \bibnamefont {Janka}}, \bibinfo {author} {\bibfnamefont
  {O.}~\bibnamefont {Just}}, \bibinfo {author} {\bibfnamefont {M.}~\bibnamefont
  {Oertel}}, \ and\ \bibinfo {author} {\bibfnamefont {N.}~\bibnamefont
  {Stergioulas}},\ }\href@noop {} {\bibfield  {journal} {\bibinfo  {journal}
  {AIP Conference Proceedings}\ }\textbf {\bibinfo {volume} {2127}},\ \bibinfo
  {pages} {020013} (\bibinfo {year} {2019})}\BibitemShut {NoStop}%
\bibitem [{\citenamefont {Fonseca}\ \emph {et~al.}(2016)\citenamefont {Fonseca}
  \emph {et~al.}}]{Fonseca:2016tux}%
  \BibitemOpen
  \bibfield  {author} {\bibinfo {author} {\bibfnamefont {E.}~\bibnamefont
  {Fonseca}} \emph {et~al.},\ }\href {\doibase 10.3847/0004-637X/832/2/167}
  {\bibfield  {journal} {\bibinfo  {journal} {Astrophys. J.}\ }\textbf
  {\bibinfo {volume} {832}},\ \bibinfo {pages} {167} (\bibinfo {year}
  {2016})},\ \Eprint {http://arxiv.org/abs/1603.00545} {arXiv:1603.00545
  [astro-ph.HE]} \BibitemShut {NoStop}%
\bibitem [{\citenamefont {Antoniadis}\ \emph {et~al.}(2013)\citenamefont
  {Antoniadis} \emph {et~al.}}]{Antoniadis:2013pzd}%
  \BibitemOpen
  \bibfield  {author} {\bibinfo {author} {\bibfnamefont {J.}~\bibnamefont
  {Antoniadis}} \emph {et~al.},\ }\href {\doibase 10.1126/science.1233232}
  {\bibfield  {journal} {\bibinfo  {journal} {Science}\ }\textbf {\bibinfo
  {volume} {340}},\ \bibinfo {pages} {6131} (\bibinfo {year} {2013})},\ \Eprint
  {http://arxiv.org/abs/1304.6875} {arXiv:1304.6875 [astro-ph.HE]} \BibitemShut
  {NoStop}%
\bibitem [{\citenamefont {Cromartie}\ \emph {et~al.}(2019)\citenamefont
  {Cromartie} \emph {et~al.}}]{NANOGrav:2019jur}%
  \BibitemOpen
  \bibfield  {author} {\bibinfo {author} {\bibfnamefont {H.~T.}\ \bibnamefont
  {Cromartie}} \emph {et~al.} (\bibinfo {collaboration} {NANOGrav}),\ }\href
  {\doibase 10.1038/s41550-019-0880-2} {\bibfield  {journal} {\bibinfo
  {journal} {Nature Astron.}\ }\textbf {\bibinfo {volume} {4}},\ \bibinfo
  {pages} {72} (\bibinfo {year} {2019})},\ \Eprint
  {http://arxiv.org/abs/1904.06759} {arXiv:1904.06759 [astro-ph.HE]}
  \BibitemShut {NoStop}%
\bibitem [{\citenamefont {Linares}\ \emph {et~al.}(2018)\citenamefont
  {Linares}, \citenamefont {Shahbaz},\ and\ \citenamefont
  {Casares}}]{Linares:2018ppq}%
  \BibitemOpen
  \bibfield  {author} {\bibinfo {author} {\bibfnamefont {M.}~\bibnamefont
  {Linares}}, \bibinfo {author} {\bibfnamefont {T.}~\bibnamefont {Shahbaz}}, \
  and\ \bibinfo {author} {\bibfnamefont {J.}~\bibnamefont {Casares}},\ }\href
  {\doibase 10.3847/1538-4357/aabde6} {\bibfield  {journal} {\bibinfo
  {journal} {Astrophys. J.}\ }\textbf {\bibinfo {volume} {859}},\ \bibinfo
  {pages} {54} (\bibinfo {year} {2018})},\ \Eprint
  {http://arxiv.org/abs/1805.08799} {arXiv:1805.08799 [astro-ph.HE]}
  \BibitemShut {NoStop}%
\end{thebibliography}%
	
\end{document}